\documentclass[prl,twocolumn,showpacs]{revtex4}
\usepackage{graphicx}
\usepackage{amsmath}
\usepackage{natbib}
\newcommand{\n}{\nonumber}
\newcommand{\bn}{\begin{eqnarray}}
\newcommand{\en}{\end{eqnarray}}
\newcommand{\eml}{\end{multline}}
\newcommand{\bml}{\begin{multline}}
\newcommand{\h}{\hspace}

\begin{document}

\title {Quantum Pumping with Ultracold Atoms on Microchips: Fermions versus Bosons}
 \author{Kunal K. Das$^1$, and Seth Aubin$^2$}
 \affiliation{
 $^1$Department of Physical Sciences, Kutztown University of Pennsylvania, Kutztown, Pennsylvania 19530, USA\\
 $^2$Department of Physics, College of William and Mary, Williamsburg, VA 23185, USA}

\date{\today }
\begin{abstract}
We present a design for simulating quantum pumping of
electrons in a mesoscopic circuit with ultra-cold atoms in a micro-magnetic  chip
trap. We calculate theoretical results for quantum
pumping of  both bosons and fermions, identifying differences and
common features, including geometric behavior and resonance
transmission. We analyze the feasibility of experiments with bosonic
$^{87}$Rb and fermionic $^{40}$K atoms with an emphasis on reliable
atomic current measurements.
\end{abstract}
\maketitle

Pumping is any cyclical time-varying mechanism  that generates sustained flow. Quantum pumping \cite{thouless,brouwer-1} in mesoscopic solid state circuits is a coherent quantum process for generating directed transport of charge with time dependent potentials, but no applied bias field. With its promise of precise and reversible flow control at the single electron level and extension to transport of spin \cite{chamon-spin} and entangled electron pairs
\cite{das-PRL}, quantum pumping  has been the
subject of considerable theoretical research
\cite{buttiker-floquet}.
Despite potential technological applications, quantum pumping experiments in solid state system have not been successful, partly due to
dominant competing rectification effects associated with
electrically charged carriers \cite{watson,switkes,brouwer-2}.
Neutral ultra-cold atomic systems present a possible path around the
current bottleneck by avoiding such complications. An atomic
circuit using a Bose-Einstein condensate (BEC) or a degenerate Fermi
gas (DFG) can test basic theoretical predictions, while also providing a
reference for experiments in solid state systems.

In this letter, we present a design for an experiment to test quantum pumping theory with ultra-cold atoms in a micro-magnetic
potential on a chip. Ultra-cold atoms open up the possibility of
studying not only fermion quantum pumping but also boson pumping, as
well as the influence of variable interactions and long range order,
in a fully controlled and tunable
system. We present theoretical results for both types of atoms in prototypical pumping schemes and we analyze the feasibility of a cold atom based experiment with numerical simulations.

\begin{figure}[b]\vspace{-8mm}
\includegraphics*[width=\columnwidth]{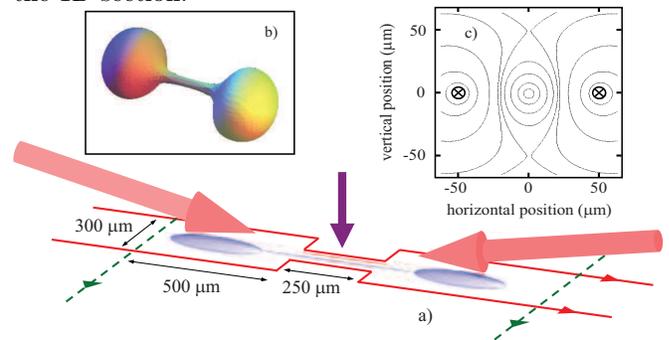}\vspace{-4mm}
\caption{(a) Configuration for generating two micro-magnetic trap
reservoirs connected by a  1-d channel. The red wires provide radial
confinement, while the dashed green `end-cap' wires, located 50 $\mu
m$ below, provide axial confinement. The large red arrows are probe
lasers for measurements on the trapped atoms, represented by the
blue structure. The vertical (purple) laser implements the pump
potential; (b) a 1 $\mu K$ equipotential for alkali atoms trapped by
250 mA and 10 mA in the red and green wires, respectively, along
with a 1 G axial magnetic field (the transverse/axial scale is 37);
(c) transverse isopotential curves along the 1-d channel from 50 to
1000 $\mu$K showing its symmetry and significant trap-depth. }
\label{Fig1}
\end{figure}

{\bf Mesoscopic circuit with atom chips:} A prototypical
mesoscopic circuit consists of a device, e.g. a quantum dot,
connected by nanowires to macroscopic contacts.  At low temperatures, electrons and holes can have mean free paths longer than the nanowires, so they can be described as freely propagating particles in one dimension (1D). The device
presents a scattering potential for the particles, so that
transport is reduced to a scattering problem \cite{imry}.

We can simulate this setup with the atom chip based scheme shown in
Fig.~\ref{Fig1}. Atom chips are substrates on which currents in
lithographically imprinted wires generate a micro-magnetic trapping
potential for ultra-cold atoms. These chips can efficiently produce
both BECs and DFGs with
temperatures in the 10 nK to 1 $\mu$K range
\cite{Hansch-chipBEC,Aubin-NaturePhysics}. Two reservoirs connected
by a 1D quantum wire can be implemented on an atom chip by two 3D
micro-magnetic traps connected by a quasi-1D magnetic guide,
generated by co-propagating currents in two parallel wires (red in
Fig.~\ref{Fig1}) on the substrate, with a constriction for the
tighter 1D section. The atoms are trapped in the plane of the wires,
with the substrate between them removed \cite{eriksson-atom-chip},
which also allows optical access from above and below.  The
trapping potential is harmonic along all principal
axes, including the axial one due to a current through the two `end
cap' wires (dashed-green in Fig.~1) below the trapping plane
\cite{Schmiedmayer}. Residual defects in the trap potential can be
suppressed by applying an AC current through the principal trapping
wires, while keeping the external axial magnetic field and
the current in the end-cap wires constant \cite{trebbia-atom-chip}.
The ``device", or scattering potential, can be realized with a
dipole laser focused onto the 1D section.

The generated atomic current can be determined
from a measurement of the momentum distribution of the particle
flow, since the average current can be written as
$J=(\hbar/m)\int|\psi(k)|^2 kdk$. Bragg spectroscopy \cite{Ketterle}
is ideally suited for measurements of the momentum distribution,
since it can be selectively applied to atoms in the 1-d channel and
combined with fluorescence imaging for high signal-to-noise
detection.  A spectroscopic flag can be attached to the kicked atoms by adding the
hyperfine splitting to the base detuning of the probe lasers, thus
changing their hyperfine level.  A large fraction of the scattered
photons can be collected by a microscope lens located a few
millimeters above the atom chip and imaged onto a high sensitivity
camera.  We calculate that roughly a hundred photons per atom can be
detected with a fluorescence pulse of a few hundred microseconds.

{\bf Theory of bosonic and fermionic pumps:} Quantum pumping has been
studied exclusively for fermions in  solid state
systems, and primarily in the adiabatic regime where the pump period
exceeds the dwell time of the carriers at the potential. With atomic experiments in mind, we extend the theory of pumping to include bosons.

 As with electrons in nanowires, the dynamics of
atoms in the central segment is quasi-1D with quasi-continuum
description along the transport direction and quantized
transverse channels (n).  The axial and transverse components can be
factorized \cite{das-crossover},
$\Psi(x,y,z,\theta;t)=\sum_n\psi_n(z,t)\phi_n(r)$, (in cylindrical
symmetry); $\int dz|\psi_n(z,t)|^2$ is the population fraction in
the $n$-th channel. Scattering influences the evolution of the axial
functions, with little effect on the transverse profiles. For weak
interactions \cite{yurovsky-review}, phase fluctuations of
degenerate bosons can be neglected in the 1D section, so the axial
dynamics has an effective description in terms of a 1D non-linear
Schr\"{o}dinger equation (NLSE)
$ {\textstyle \frac{1}{2}}\left[-\partial_z^2+ (z/\gamma)^2
\right]\psi+g_{1D}|\psi|^{2}\psi= i\partial_t\psi\label{1D-eq}$
where $\gamma=\omega_r/\omega_z$ is the aspect ratio. Fermions and
non-interacting bosons are both described by setting
$g_{1D}=0$. The axial potential variation is small over the 1D
segment, so we set $z^2/\gamma^2\simeq0$, allowing a plane wave
description. The radial trap frequency $\omega_r$ sets the scale for
our expressions: the energy, length and time units are
$\hbar\omega_r$,  $l_r$=$\sqrt{\hbar/(m\omega_r)}$ and
$\omega_r^{-1}$. Time evolution is governed by the axial energy of available channels $E$=$\mu-n\hbar\omega_r$, where
$\mu$ is the chemical potential .

For a slowly varying external potential $V(x, t)$, an adiabatic
expansion to first order in the time-derivative approximates the
time-dependent scattering states
\bn\psi_{k, n}(x,t)\simeq\psi_{k,n}^t(x)-i{\textstyle \int} dx'
G^t(x,x';E)\partial_t\psi_{k,n}^t(x')\label{wf-lin-ord},\en
in terms of the solutions of the Lippmann-Schwinger (LS) equation
$|\psi_{k,n}^t\rangle=|\phi_{k,n}\rangle+\hat{G}^t(E)\hat{V}^t|\phi_{k,n}\rangle$
for the \emph{instantaneous} potential $V^t(x)$. Here $
\hat{G}^t(E)=[E-\hat{H}_0-\hat{V}^{t}]^{-1}$ is the instantaneous
Green's operator and $|\phi_{k,n}\rangle$ are scattering states of
the time-independent Hamiltonian $\hat{H}_0$, and $k,n$ label
wavevector and channel. The zeroeth order current vanishes for a
pump with no bias field  \cite{das-PRL}. Denoting the second term in
Eq.~(\ref{wf-lin-ord}) by $\Delta\psi_{k,m}(x,t)$ , the adiabatic
pumped current of spin polarized fermions is
\bn \h{-2mm}&&\h{-2mm}J_F(x,t)={\textstyle\h{-1mm}\sum_n\int dE f(E)
\int\h{-1mm}\frac{dk}{2\pi}\  \delta ({\textstyle\frac{k^2}{2}}-E)
\times}\\\h{-2mm}&&\h{4mm}{\rm Im}\left\{ \psi_{k,n}^{t*}(x)\partial_{x}
\Delta\psi_{k,n}(x,t)+\Delta\psi_{k,n}^*(x,t)\partial_{x}\psi_{k,n}^t(x)
\right\}\h{2mm}\label{1st-order-1}\n\\\h{-2mm}&&\h{-2mm}
=\h{-1mm}{\textstyle\frac{1}{2\pi}\int_0^\infty\h{-1mm}
dEf(E)\partial_E\h{-1mm}\int \h{-1mm}dx'\dot{V}(x',t) {\rm
Im}\{G_E^{t*}(x'\!,x )
\partial_{x}G_E^t(x,x')\!\}}\n\en
using Green's function identities and the LS equation
\cite{das-PRL}.   At low temperatures, the Fermi distribution is
approximately a step function, $f(E)\sim\theta(E_F-E)$, so we
obtain\bn J_F(x,t)\simeq \int\frac{dx'}{2\pi}
\dot{V}(x'){\rm Im}\{G_{E_F}^*(x',x)
\partial_{x}G_{E_F}(x,x')\}\label{adiabatic-fermions}\en

In order to describe bosons, we use the parabolic dispersion of
plane waves to replace $\int_0^\infty \!dE
f(E)\frac{\partial}{\partial E}\rightarrow\int_{0}^{\infty} dk
f({\textstyle \frac{k^2}{2}})\frac{\partial}{\partial
k}\stackrel{T\rightarrow0}{\rightarrow}\int_{0}^{k} dk
\frac{\partial}{\partial k}$ . For non-interacting degenerate bosons
at rest or in a superposition of  momentum states, $|k\rangle$ and
$|-k\rangle$, the pumped current is
\bn J_B(x,t)={\textstyle\frac{\partial}{\partial k}\int}
dx'\dot{V}(x'){\rm Im}\{G_E^*(x'\!,x)
\partial_{x}G_E(x,x')\!\}\label{adiabatic-bosons}\en
Thus the pumped current for a degenerate fermi gas at fermi-vector
$k$ is related to the pumped current for degenerate bosons at
wavevector $k$ by $ J_B(k)=2\pi\frac{\partial}{\partial k}
J_F(k^2/2)$.  For a BEC at rest, one sets $k=0$, after the
derivative.

Essential features of quantum pumps can be understood with models involving time-varying \emph{single barrier} potentials: with variable
strength $ V_1(x,t)=U(t)\delta(x)$ or translating uniformly
$V_2(x,t)= U\delta(x-vt)$.
For \emph{adiabatic} variation, Eqs.~(\ref{adiabatic-fermions}) and
(\ref{adiabatic-bosons}) give the pumped currents:
{\small \bn &&\h{-2mm}J_F^{(1)}(x,t)={\rm sgn(x)}\frac{1}{2
\pi}\frac{\dot{U}k_F}{k_F^2+U^2};\h{2mm}
J_F^{(2)}=\frac{1}{\pi}\frac{k_F v
U^2}{k_F^2+U^2}\label{ad-limit}\\
&& \h{-2mm}J_B^{(1)}(x,t)={\rm
sgn(x)}\frac{\dot{U}(U^2-k^2)}{(U^2+k^2)^2};\h{2mm}
J_B^{(2)}=\frac{2vU^2(U^2-k^2)}{(U^2+k^2)^2}\n\en}
They show the role of symmetry:  $V_1$ generates no net particle
transport from one reservoir to the other over a period, due to antisymmetry with
respect to $x$; while $V_2$ being symmetric leads to net transport.
In general,  both symmetric and antisymmetric
parts can be present.

\begin{figure}[t]
\includegraphics*[width=\columnwidth]{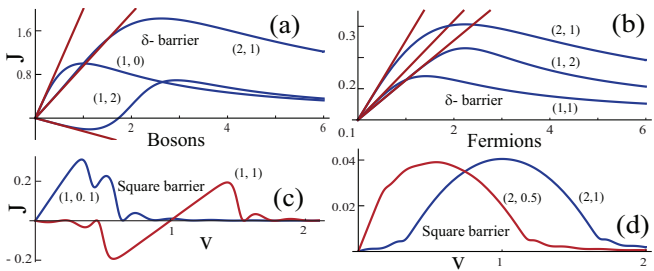}\vspace{-3mm}
\caption{Pumped current versus velocity of a uniformly
translating potential: (a,b) $\delta$-barrier, (straight lines
are adiabatic approximations); (c,d) square barrier of width $a=10\ l_r$ and height $U/a$. Barrier strengths and wavevectors, $(U,k)$ for bosons
and $(U,k_F)$ for fermions, are shown. } \label{Fig-1-delta}\vspace{-6mm}
\end{figure}

The fermionic current for $V_2$ is always in the direction of motion
of the potential, but the bosonic current can flow opposite
( Fig.~\ref{Fig-1-delta}): When the bosons have sufficient energy $k^2>U$,
the transmitted fraction dominates, and particles
going against the barrier have a higher transmission probability;
for fermions, the averaging over states washes out this effect. Over
a period $T$, the net pumped particles, $J\times T$, is independent of the velocity $v$ and depends only on the parameter path traversed by the potential. It is a \emph{geometric} quantity  analogous to a geometric phase
\cite{zhou-mckenzie} , a feature
shared by all adiabatic quantum pumps.

The pumped current at \emph{arbitrary} barrier velocities for $V_2$
can be found using a Galilean transformation to be: {\small\bn
J_B\!=\!\frac{2v U^2  (U^2 \!+ \!v^2-k^2 )}{U^4\! \! +\! 2 U^2
(v^2\! \!+ \!k^2)\!+\! (v^2 \!\!- k^2)^2};\h{1mm}
J_F\!=\!\frac{U^2}{4\pi}\!\!\ln\h{-1mm}\left|\frac{U^2\!\!+\!(k_F\!+\!v)^2}
{U^2\!\!+\!(k_F\!-\!v)^2}\right|\en}
The adiabatic expressions in Eq.~(\ref{ad-limit}) are retrieved with a
Taylor expansion for $v\ll k, k_F$. The fermionic current
vanishes for $k_F$=$0$, as the number density vanishes; but for a stationary BEC, $J_B(k$=$0)=vU^2/(v^2+U^2)$, entirely due to reflection.
At high barrier velocity $v\gg k$, $J_B\sim J_B(k$=$0)$ and $J_F\sim k_FvU^2/(\pi
(v^2+U^2))$ like the adiabatic limit with $k_F$ and $v$
interchanged.

Even for a \emph{finite} translating square barrier (SB)
$V_{SB}=(U/a)\theta(x-vt)\theta(-x+vt+a)$ of width $a$ and height
$U/a$, analytical expressions for the pumped current can be likewise
calculated, too lengthy to be shown, but plots based upon those
solutions are shown in Fig.~\ref{Fig-1-delta}.  The
pumped current differs dramatically from the case of the delta
barrier: (i) the finite height allows particle energy to exceed the
barrier potential leading to sharp transitions at $\frac{1}{2}(k\pm
v)^2=U$, the classical cutoffs for transmission, and (ii) the finite
width creates oscillations due to resonance transmission. For bosons
(Fig.~\ref{Fig-1-delta}(c)), the oscillations are pronounced, with
the current vanishing and reversing for some velocities; but less so
for fermions (Fig.~\ref{Fig-1-delta}(d))  due to averaging over
wavevectors. For a translating \emph{barrier}, both classical and quantum features are manifest, but for a translating \emph{well}, the behavior is quantum mechanical.

The quantum nature of pumping becomes truly significant in a \emph{turnstile pump} comprised of two barriers with heights oscillating out of phase with each other. This model has been studied \cite{buttiker-floquet,das-opatrny} for
fermions, and here we present results for bosons contrasted with
fermions. Essential features can be understood with two delta
function potentials $U_\pm(t) \delta(x\mp a)$ with oscillating
strengths $U_+(t)$=$1+\cos(\omega t)$ and $U_-(t)$=$1+\sin(\omega t)$,
that trace out a circle over a period
$T$=$2\pi/\omega$. In this limiting case, the current is entirely due to quantum interference \cite{das-opatrny}. Reversing the cycle reverses the flow.

\begin{figure}[t]
\includegraphics*[width=\columnwidth]{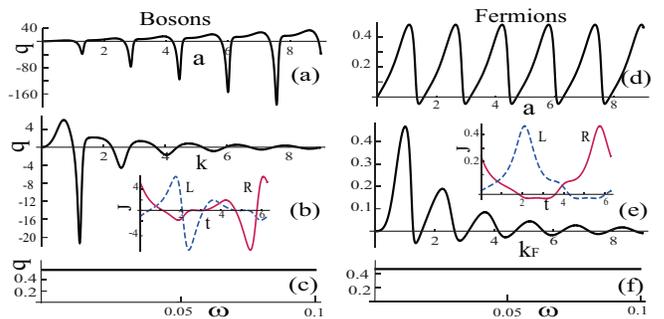}\vspace{-4mm}
\caption{A double barrier pump of bosons (a)-(c) and
fermions (d)-(f). Particle transport
in a pump cycle as a function of: (a,d) barrier separation, 2a, at
 \emph{k=k$_F$=}1; (b,e) wavevector k (fermi-vector k${\rm_F}$ for fermions)
at \emph{a=}1; (c,f) angular frequency $\omega$, at \emph{a=k=k$_F$=}1. Insets show current versus time at the right and left reservoirs . The
length and time units are $l_r$ and
$\omega_r^{-1}$.}\label{Figure-Double-Delta}\vspace{-5mm}
\end{figure}

The insets in Figure~\ref{Figure-Double-Delta} show that the
currents on the left and the right of the potential are not in sync
and vary over time, but their time integrals over a full cycle are
equal. Resonant transmission effects are prominent due to finite
barrier separation, 2a: The particle transport, \emph{q} , in a pump
cycle displays oscillations and peaks as a function of the barrier
separation, Fig.~\ref{Figure-Double-Delta}(a,d), and also as a
function of the wavevector $k$ or $k_F$,
Fig.~\ref{Figure-Double-Delta}(b,e). Fermions display less
pronounced resonance behavior, due to averaging over
momentum states. There has been recent interests in testing
resonance transport through double barrier structures
\cite{paul-resonant-PRL}, quantum pumps  demonstrate this by
periodic cycling of the potentials. The geometric nature of
adiabatic pumps is clearly seen in
Fig.~\ref{Figure-Double-Delta}(c,f), since particle transport
per cycle is independent of $\omega$.

{\bf Feasibility Analysis:} Pump potentials can be implemented with blue-detuned lasers at 532 nm
focused to 1-5 $\mu$m  gaussian-profile barriers.
The lasers need to translate at velocities $v\sim
l_r\omega_r\simeq 0.5$ cm/s or vary in intensity at frequencies
$\sim\omega_r\simeq 2$ kHz, easily achievable.

\begin{figure}[t]
\includegraphics*[width=\columnwidth]{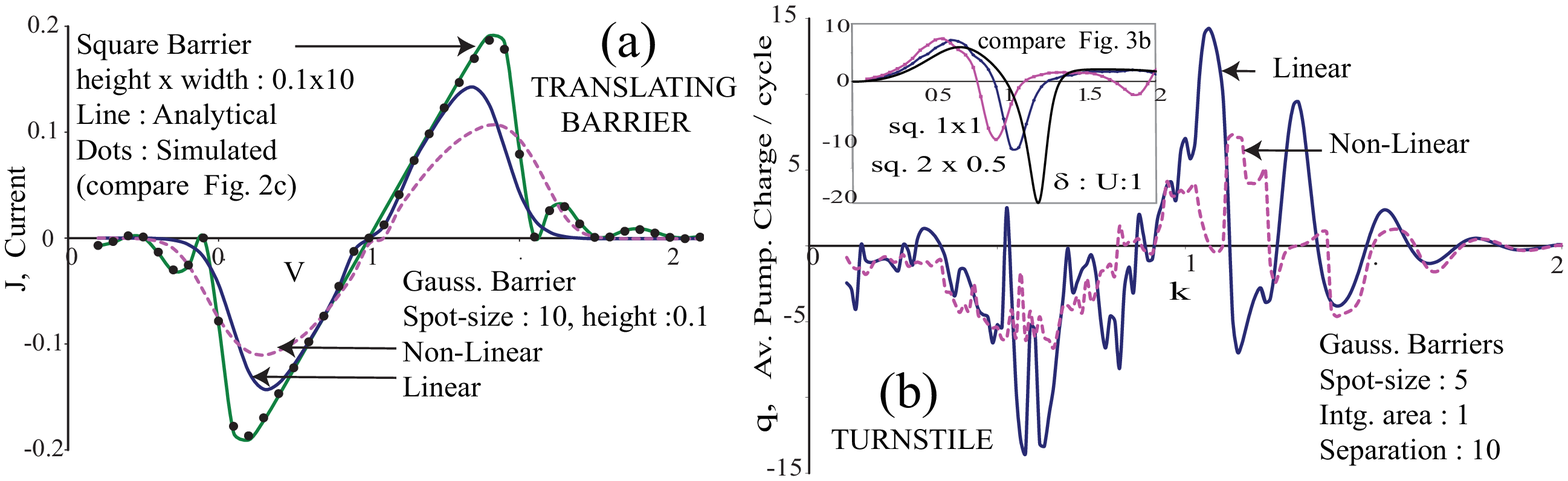}\vspace{-4mm}
\caption{Numerical simulation of pumping with interacting and non-interacting bosons. In (a) k=1. Square barrier (SB) plots
show consistency with analytical results. The inset compares simulations for a SB turnstile of separation 2a=2 \emph{with} analytical results for a $\delta-$barrier turnstile as in Fig. 3(b). As the barrier width decreases, with area fixed at 1, the SB results approach the $\delta-$barrier analytical results.
}\label{Figure-simulations}\vspace{-5mm}
\end{figure}

\emph{Bosonic pumps} at non-zero $|k|$ can
be implemented with a broad (relative to pump potential) wavepacket
split into counterpropagating momentum states by a Bragg pulse \cite{Ketterle}. In
this scenario, the reservoirs can be removed. For $^{87}Rb$ in the
$F=2$, $m_f=+2$ state in the set-up of Fig. \ref{Fig1}, the
transverse and axial trap frequencies are $\omega_{r,1D}=2\pi\times
5.1$ kHz and  $\omega_{axial}=2\pi\times 3.6$ Hz. For a wavepacket
of 1000 $^{87}Rb$ atoms with scattering length of $a_s=99 a_0$ a
variational calculation \cite{das-anisotropic} yields the
effective 1D non-linear constant $g_{1D}=67.3$ and axial
Thomas-Fermi width $587\ l_r$. After the axial trap is turned off
and the Bragg pulse applied, the split wave packet evolves in the presence of the pump potential. We simulate this numerically by solving
the NLSE with a split-step operator method; results are
shown in Fig.~4. In the absence of nonlinearity ($g_{1d}=0$),
the wavepacket simulations with \emph{square} barriers are
consistent with analytical results (Figs. 2(c) and 3(b)) obtained assuming
plane waves, validating the method.  Figure \ref{Figure-simulations} also
shows that Gaussian profile barriers
and  nonlinearity lead to some qualitative changes,
but the pumped current or charge remains significant. The
nonlinearity reduces the signal somewhat, and for the turnstile, the
broader barriers and barrier-separation lead to more closely
spaced oscillations in the current as packet velocity (k) varies.
Numerical simulations \cite{das-wavepacket} show that the pump signal for the \emph{turnstile} is more sensitive to chip trap roughness than the translating barrier scheme, but AC suppression of roughness \cite{trebbia-atom-chip} is sufficient for a robust signal.

A 1000 atom wavepacket has initial peak density $7.3\times
10^{14}\ cm^{-3}$ and chemical potential $\mu_{3d}=0.26\mu K =
1.09\hbar\omega_r$. The number of atoms can be significantly
increased, considering: (i) More atoms mean stronger
nonlinearity and faster expansion, requiring longer traps to allow
sufficient interaction times with the pump; without the reservoirs
the axial length can be extended upto $1000\ \mu m$. (ii) To remain
in the transverse ground state (for single channel),
$\mu<2\hbar\omega_r$;  our variational calculations gives $\mu\simeq
1.6\hbar\omega_r$ with  $N_{1D}=2.0\times 10^4$ atoms.

For \emph{fermion pumps} with $^{40}K$ the currents listed
for Fig. \ref{Fig1} produce trap frequencies a factor of
$\sqrt{m_{Rb}/m_{K}}\simeq1.5$ higher than with $^{87}Rb$.
Energetically, the 1D section can contain
$\omega_{r,1D}/\omega_{axial}\simeq1400$ spin-polarized fermions in
the lowest transverse channel due to the Pauli principle. Since the
size of harmonic oscillator eigenstates scales as
$\sqrt{2N}$, for the axial oscillator length $l_z=6.9\ \mu m$, the
$250\ \mu m$  1D section will hold about 700 atoms; each reservoir
contains 50 times more. The lowest channel can
accommodate Fermi vectors up to $k_F=1.4 l_r^{-1}$.

\emph{Conclusions}: Our analysis has shown that quantum pumping experiments can be done with current atom-chip technology,
allowing a broad survey of a process that has eluded
confirmation in solid state systems. In addition to simulating
fermion pumping, ultra-cold atom based experiments open up
the possibility of studying quantum pumping of bosons which we
expect to show enhanced resonant tunneling and
current reversal.  Furthermore, the scheme can be adapted to search for conductance quantization for pumping with periodic lattices \cite{thouless} by imposing a moving optical  lattice on the 1-d quantum channel. In a broader context, our design
is easily adapted to a variety of mesocopic transport experiments,
important in electronic systems, like conductance quantization and
spin
transport, yet hardly explored with ultracold atoms.\\

K.K.D. acknowledges support of the Research Corporation in the initial stages.

\end{document}